\documentclass[conference, letterpaper]{IEEEtran}
\IEEEoverridecommandlockouts
\usepackage{cite}
\usepackage{amsmath,amssymb,amsfonts}
\usepackage{algorithmic}
\usepackage{graphicx}
\usepackage{textcomp}
\usepackage{booktabs}
\usepackage{url}
\usepackage{multirow}
\usepackage{xcolor}
\usepackage{listings}
\usepackage{hyperref}

\colorlet{punct}{red!60!black}
\definecolor{background}{HTML}{F5F5F5}
\definecolor{delim}{RGB}{20,105,176}
\colorlet{numb}{magenta!60!black}
\definecolor{stringcolor}{RGB}{0,128,0}

\lstdefinelanguage{json}{
    basicstyle=\small\ttfamily,  
    numbers=left,
    numberstyle=\tiny\color{gray},  
    stepnumber=1,
    numbersep=8pt,
    showstringspaces=false,
    breaklines=true,
    frame=single,  
    backgroundcolor=\color{background},
    stringstyle=\color{stringcolor},  
    commentstyle=\color{gray}\itshape,
    morestring=[b]",
    literate=
     *{0}{{{\color{numb}0}}}{1}
      {1}{{{\color{numb}1}}}{1}
      {2}{{{\color{numb}2}}}{1}
      {3}{{{\color{numb}3}}}{1}
      {4}{{{\color{numb}4}}}{1}
      {5}{{{\color{numb}5}}}{1}
      {6}{{{\color{numb}6}}}{1}
      {7}{{{\color{numb}7}}}{1}
      {8}{{{\color{numb}8}}}{1}
      {9}{{{\color{numb}9}}}{1}
      {:}{{{\color{punct}{:}}}}{1}
      {,}{{{\color{punct}{,}}}}{1}
      {\{}{{{\color{delim}{\{}}}}{1}
      {\}}{{{\color{delim}{\}}}}}{1}
      {[}{{{\color{delim}{[}}}}{1}
      {]}{{{\color{delim}{]}}}}{1},
}

\def\BibTeX{{\rm B\kern-.05em{\sc i\kern-.025em b}\kern-.08em
    T\kern-.1667em\lower.7ex\hbox{E}\kern-.125emX}}

\begin{document}
\bstctlcite{IEEEexample:BSTcontrol}
\title{Automatic Association of Cloud Security Controls and Quantifiable Metrics for Certification
\thanks{This work is partially supported by EMERALD - Evidence Management for Continuous Certification as a Service in the Cloud (101120688) under the Horizon Europe program funded by the EU; by SERICS (PE00000014) under the NRRP MUR program funded by the EU - NGEU.}
}

\author{\IEEEauthorblockN{John Bianchi\IEEEauthorrefmark{1}\IEEEauthorrefmark{2}, Shuya Dong\IEEEauthorrefmark{3}, Luca Petrillo\IEEEauthorrefmark{1}\IEEEauthorrefmark{2} and Marinella Petrocchi\IEEEauthorrefmark{1}\IEEEauthorrefmark{2}}
\IEEEauthorblockA{\IEEEauthorrefmark{1}\textit{IMT School for Advanced Studies Lucca, Lucca, Italy} \\
\IEEEauthorrefmark{2}\textit{Institute for Informatics and Telematics (IIT-CNR), Pisa, Italy} \\
\IEEEauthorrefmark{3}\textit{Independent Researcher} \\
Emails: \{john.bianchi, luca.petrillo\}@imtlucca.it, marinella.petrocchi@iit.cnr.it}
}

\maketitle
\begin{abstract}
The draft candidate European Cybersecurity Certification Scheme for Cloud Services (EUCS) defines security controls that must be associated with measurable metrics to assess compliance. This association process is currently manual, time-consuming, and prone to inconsistencies.
In this paper, we propose an automated approach based on Sentence Transformers to associate cloud security controls with quantifiable metrics by leveraging semantic similarity between their textual descriptions. We evaluate our method on a dataset of 70 controls derived from the EUCS framework.
The proposed approach outperforms a FastText-based baseline, achieving a conditional Normalized Discounted Cumulative Gain at rank 10 score of 0.640 (+0.146) and improving the standard nDCG@10 score from 0.275 to 0.504. These results demonstrate that contextual embedding models significantly enhance both the likelihood of retrieving relevant metrics and their ranking quality.
Our findings highlight the potential of transformer-based methods to support automated, scalable, and more reliable compliance processes in cloud cybersecurity certification.
\end{abstract}


\section{Introduction}
The rapid expansion of cloud computing has brought significant benefits in terms of scalability, cost-efficiency, and accessibility. However, this growth has also introduced complex cybersecurity challenges that must be addressed to ensure the safety and privacy of data in the cloud. Recognizing the importance of securing cloud services, regulatory bodies across the globe have defined certification schemes to enforce robust security standards. One such initiative is the draft candidate European Cybersecurity Certification Scheme for Cloud Services (EUCS)\footnote{\url{https://www.enisa.europa.eu/publications/eucs-cloud-service-scheme} ENISA, “EUCS – Cloud Services Scheme,” Draft version \cite{ENISA2020}. All URLs were last accessed on March 24, 2026.}, which is part of the broader framework established under the EU Cybersecurity Act\footnote{\url{https://digital-strategy.ec.europa.eu/en/policies/cybersecurity-act}}. The EUCS aims to harmonize cybersecurity practices across Europe, providing a structured approach for assessing and certifying the security of cloud services.

For cloud providers, complying with these regulatory controls is crucial not only to gain certification but also to maintain trust with customers and stakeholders. However, part of the process of demonstrating compliance with controls dictated by a certification scheme is still manual, time-consuming, and prone to errors. 

The traditional approach requires cloud providers to i) manually map controls in natural language to specific quantifiable metrics, ii) assess whether these metrics are met, and iii) generate the necessary documentation for certification. The manual mapping poses several challenges, including the difficulty of staying updated with evolving regulations, the potential for human error in interpretation, and the inefficiency of repetitive tasks.

Automating the association between controls, such as those specified by the EUCS, and the corresponding metrics for measuring compliance offers a promising solution to these challenges. Automation can streamline the compliance process, ensuring that cloud providers can quickly and accurately demonstrate adherence to regulatory standards. By leveraging automated tools, cloud providers can reduce the time and resources required for compliance, minimize the risk of human error, and enhance their ability to respond to regulatory changes. Moreover, automation can provide continuous monitoring and real-time reporting, enabling cloud providers to maintain a high level of security assurance and readiness for certification audits.

This paper explores the advantages of automating the association between regulatory controls and quantifiable metrics in the context of cloud cybersecurity certification. In particular, we consider part of the security controls of the EUCS draft version 2020 \cite{ENISA2020} and all the 163 metrics defined and associated with them by the partners of the European project MEDINA, which ended in October 2023\footnote{MEDINA: Security framework to achieve a continuous audit-based certification in compliance with the EU-wide cloud security certification scheme \url{https://medina-project.eu/}}.

By adopting a sentence transformer-based approach, we improve upon a previous method based on feature vector representations and clustering. The model reaches a conditional nDCG@10 (Normalized Discounted Cumulative Gain at 10) of 0.640. This value is computed only for controls where at least one relevant metric appears in the top-10. It is 0.146 points above the FastText baseline. Across all 70 controls, the average nDCG@10 improves from 0.275 to 0.504.

\section{The European Cybersecurity Certification Scheme for Cloud Services}

A draft version of the EUCS \cite{ENISA2020}, aimed at certifying cloud services with respect to cybersecurity guarantees, was made available for public review in December 2020.

The EUCS is the second certification scheme developed under the framework of the EU Cybersecurity Act (EU CSA) \cite{EUCSA}. In response to a request from the European Commission in November 2020, ENISA formed an ad-hoc working group representing various stakeholders (such as industry professionals, auditors, and academia) to aid in developing the candidate certification scheme for cloud services.

The EUCS is informed by several key sources, including the recommendations from the expert group CSPCERT \cite{CSPcert}, BSI C5 \cite{BSIc5}, SecNumCloud \cite{secnumcloud}, and various ISO standards. 

The EUCS adheres to the three security levels outlined in Article 52(6) of the EU CSA, which are ‘basic’, ‘substantial’, and ‘high’. In this framework, the cloud services’ security controls and the corresponding assessments increase in complexity across these levels, in terms of scope, rigor, and depth. According to the EUCS draft candidate version, \textit{the controls at level ‘high’ are demanding and align closely with state-of-the-art practices, while those at level ‘basic’ establish a minimum acceptable standard for cloud cybersecurity}.

\subsection{Security controls and security metrics}
In IT security, controls and metrics are crucial in establishing and maintaining robust security schemes. Security controls refer to the conditions that must be fulfilled for a process to achieve its security objectives. Metrics provide quantifiable measures to evaluate if the conditions expressed in the controls are fulfilled or not. 
For example, as shown in Figure \ref{fig:metrics-reqs}, the EUCS control OPS-05.3H on protection against malware is paired in MEDINA with the metric MalwareProtectionEnabled, which checks whether the antimalware solution is active.
Metrics can be specified over a period, such as six months or a year, and they are frequently expressed as percentages, numbers (integers or reals), booleans, or reference values inside an interval (scales). They also often have formulas. When using continuous monitoring techniques, it is crucial to specify how frequently a metric value is collected. Measurements must be easily obtained, and the method must be repeatable, consistent, and reproducible for them to be relevant. 

Metrics provide a quantitative basis for assessing whether the conditions expressed by a security control are fulfilled; it is therefore essential to link the right metrics to the right controls. In the MEDINA catalogue these metrics are defined as operational, tool-measurable indicators rather than direct measures of residual risk. The association allows for informed decision-making regarding resource allocation and prioritization of security initiatives. Additionally, many industries are subject to regulatory controls that mandate specific security measures. Metrics help demonstrate compliance with these regulations, ensuring accountability and transparency in security practices. 

\subsection{From manual to automatic association}\label{sec:linkingMetricsAndControls}

\begin{figure*}[t]
    \centering
    \includegraphics[width=0.9\linewidth]{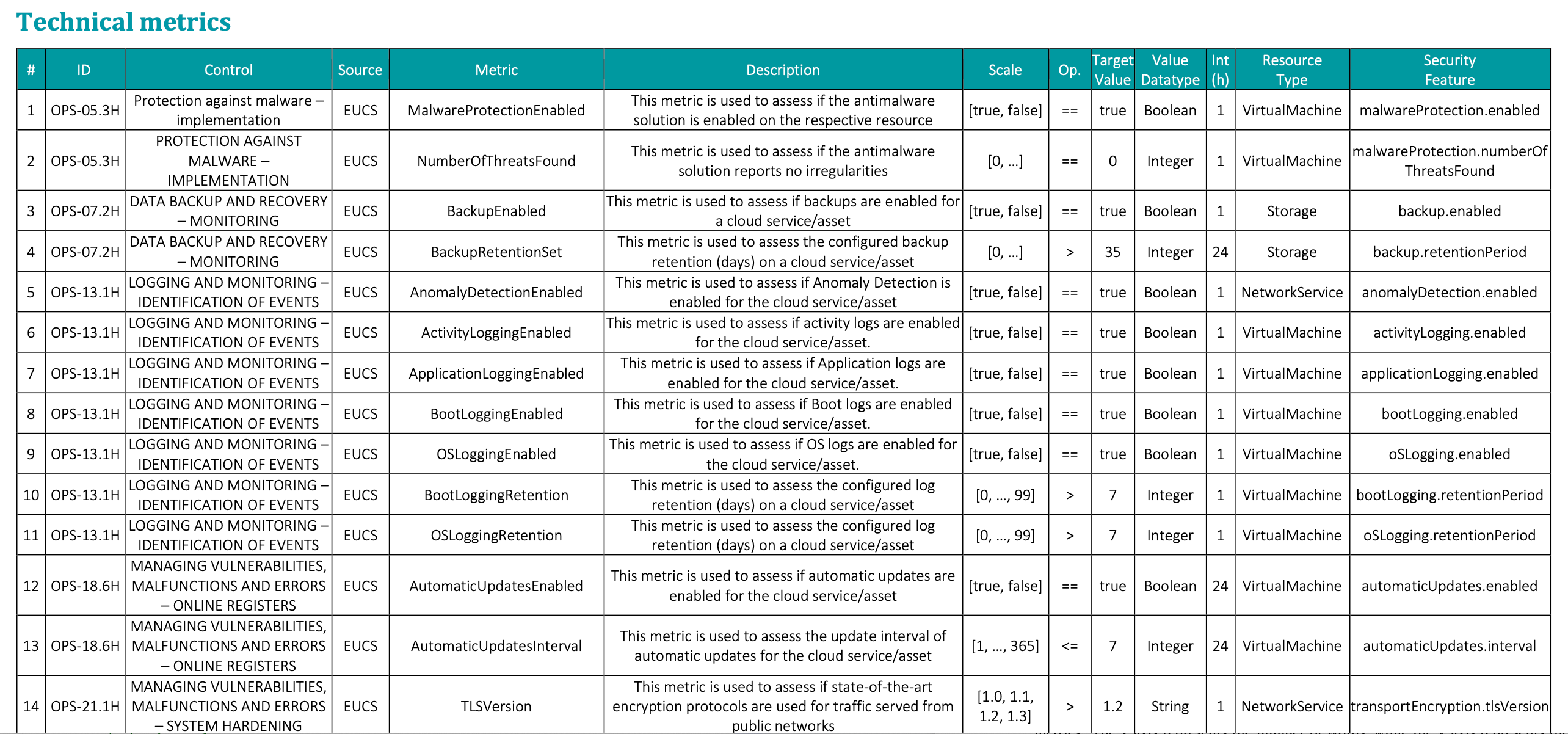}
    \caption{Examples of associations between controls and metrics in EUCS (Property of the MEDINA consortium, Deliverable 2.2 \cite{medinad2})}
    \label{fig:metrics-reqs}
\end{figure*}

As anticipated in the introduction, a manual mapping of a subset of the EUCS controls to a set of metrics was performed by the MEDINA consortium. The associations, documented in the Appendix of Deliverable 2.2 \cite{medinad2}, result from an iterative expert review involving multiple partners (including TECNALIA\footnote{\url{https://www.tecnalia.com/en/}}, CNR\footnote{\url{https://www.cnr.it/en}}, Bosch\footnote{\url{https://www.bosch.com/}}, Fraunhofer\footnote{\url{https://www.fraunhofer.de/en.html}}, Fabasoft\footnote{\url{https://www.fabasoft.com/}}, HPE\footnote{\url{https://www.hpe.com/}} and XLAB\footnote{\url{https://www.xlab.si/}}), followed by internal quality review and approval by all partners. Metrics were elicited from established sources such as NIST SP 800-55\footnote{\url{https://csrc.nist.gov/pubs/sp/800/55/r1/final}} and prior EU projects, and were refined in a second iteration to retain only actionable, tool-measurable definitions. Figure \ref{fig:metrics-reqs} shows an excerpt of these consensus-based associations.
The Metric column shows the name given to the metric, the Description column gives its textual description, and the ID column is the identifier of the control to which the metric is associated. The Control column shows the control category.
MEDINA lists only "antimalware enabled" as relevant, so in an ideal ranking our model would place it first, while a clearly unrelated metric such as "percentage of successful backups" would appear beyond position 10.

Since manual association is challenging and time-consuming, the ideal way to proceed would be to obtain the associations automatically. 
In the rest of this article, we take the associations made in MEDINA as ground truth and proceed to test automatic methods for obtaining them.

\section{Dataset}\label{sec:dataset}
The MEDINA project partners have defined in a textual format the quantifiable metrics for evaluating a subset of the EUCS system controls. Specifically, 163 metrics were defined and manually mapped to 70 controls (in fact, multiple metrics can be associated with one control).
In the following, we present experiments that we conducted to automatically associate these 163 metrics with the 70 controls. 

As we can see in figure \ref{fig:association_example}, both the controls and the metrics have a textual description.

\begin{figure}[htbp]
  \centering
  \begin{lstlisting}[language=json]
{
  metric: {
    description: "This metric is used to assess if the antimalware solution is enabled on the respective resource.",
    category: "Operational security",
    targetResourceType: "VirtualMachine"
  },
  control: {
    description: "The CSP shall automatically monitor the systems covered by the malware protection and the configuration of the corresponding mechanisms to guarantee fulfillment of above controls, and the antimalware scans to track detected malware or irregularities.",
    type: "Organizational"
  }
}
  \end{lstlisting}
  \caption{Textual description of a control and that of the associated metric}
  \label{fig:association_example}
\end{figure}

\begin{figure}[htb!]
  \centering
  \includegraphics[width=0.5\textwidth]{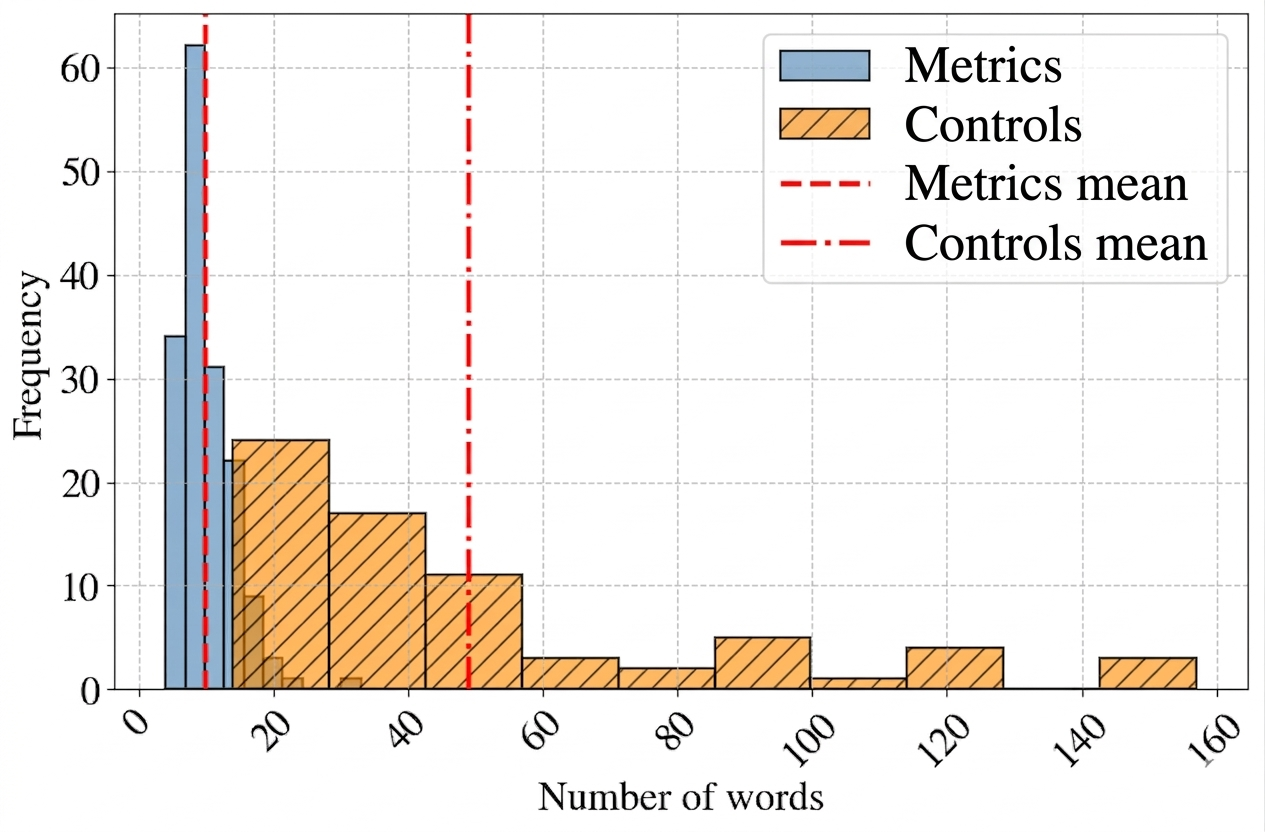}
  \caption{Distribution of the number of words in the textual description of  controls and metrics considered in this study}
  \label{img:wordDistribution}
\end{figure}

Figure \ref{img:wordDistribution} shows the distribution of word counts for the controls and the metrics in our dataset. The x-axis represents the number of words, while the y-axis represents the frequency of occurrence. Both distributions show a skewed pattern, indicating that the majority of the textual descriptions have a relatively small number of words, with fewer instances having a larger word count. The mean word count for the controls is significantly higher than that for the metrics, as shown by the vertical lines representing the means. This disparity indicates that the descriptions in the former tend to be more verbose than those in the latter. This is because the metric needs to be described as technically as possible to facilitate its implementation.

\section{From FastText to Sentence Transformers}
\subsection{Metric Recommender: The MEDINA Approach}\label{sec:oldMethod}
One of the tools developed within the MEDINA project is the Metric Recommender. It takes a natural language description of an EUCS control and a list of metrics as input and returns the association of the metrics with the control in descending order of relevance as output. 
The idea is to associate metrics with controls based on the similarity of their textual descriptions. This approach involves feature extraction and clustering steps. The textual descriptions of the controls and metrics are first transformed into feature vectors using pre-trained NLP models, and then clustered.
The feature extractor chosen is FastText\footnote{\url{https://fasttext.cc/}}, which is pre-trained on English texts from Wikipedia and Common Crawl. Data cleaning, such as removing stop words, is performed before feature vector computation to eliminate irrelevant information.  For the Sentence Transformer models, no additional preprocessing was applied beyond the models' built-in tokenization, in order to preserve domain-specific terminology and structure.
Assuming that metrics similar to a control are close to it in the feature space, a K-d tree is constructed based on the feature vectors of these metrics. This can then be used to select the k-closest neighbors of a query vector — the feature vector of a control — based on the shortest Euclidean distance.
We will test this approach on our data and present the results in Section \ref{sec:results}.

\subsection{The EMERALD Approach}
In line with one of the goals of the European project Emerald\footnote{Emerald: Evidence Management for Continuous Certification as a Service in the Cloud \url{https://www.emerald-he.eu/}}, the MEDINA's follow-up, we now try to improve the performance of the MEDINA's Metric Recommender. 
For a direct comparison with the MEDINA approach, we use the same ground truth dataset: manually made associations between 70 controls and 163 metrics.

Our proposal is still based on the idea of exploiting the similarity between texts: the more similar the metric description and the control description are, the more likely it is that the metric is associated with the control.
This task is related to Semantic Textual Similarity evaluation and has been effectively addressed by transformer models. BERT \cite{DBLP:journals/corr/abs-1810-04805}, for instance, has significantly improved the accuracy of semantic similarity calculations compared to earlier techniques \cite{chen2022semeval, yang2020measurement}. In this work, we utilize Sentence Transformers, which deliver performance comparable to traditional transformers \cite{sun2022sentence} but are optimized for speed, especially in sentence similarity tasks.

Introduced by Reimers and Gurevych in 2019, SBERT \cite{reimers2019sentence} reduces the time required to compute sentence similarity from 65 hours (when using BERT) to just a few seconds thanks to its siamese network structure. Since then, this family of models has expanded to cover various domains, with multiple variants currently available on the Hugging Face platform \cite{huggingFaceSentenceTransformers}.

In the absence of pre-trained or domain-specific options for cloud security controls and metrics, we selected the five with the highest performance across general tasks, as evaluated over 14 domains for sentence embeddings, as of March 2026 according to the Sentence Transformers website \cite{sentenceTransformersChart}. To compute similarity between embeddings, we employ cosine similarity using the scikit-learn \textit{cosine\_similarity} function\footnote{\url{https://scikit-learn.org/stable/modules/generated/sklearn.metrics.pairwise.cosine_similarity.html}}. Table \ref{tab:model_performance} provides details on the selected models, including their base architectures and the short names used in the graphs of Results (\ref{sec:results}). All five models are Sentence Transformer models trained with the SBERT siamese architecture, differing only in their underlying transformer backbone.

\begin{table}[h!]
\centering
\renewcommand{\arraystretch}{1.6} 
\caption{Sentence Transformer models for generating Sentence Embeddings}
\begin{tabular}{p{2.5cm} p{2.5cm} p{2.5cm}}
\textbf{Model Name} & \textbf{Base Model} & \textbf{Short Name} \\
\midrule
all-mpnet-base-v2 \cite{all-mpnet-base-v2} & microsoft/mpnet-base \cite{microsoft/mpnet-base} & AMPNet \\
multi-qa-mpnet-base-dot-v1 \cite{multi-qa-mpnet-base-dot-v1} & microsoft/mpnet-base & MMPNet \\
all-distilroberta-v1 \cite{all-distilroberta-v1} & distilroberta-base \cite{distilroberta-base} & DRoBERTa \\
all-MiniLM-L12-v2 \cite{all-MiniLM-L12-v2} & microsoft/MiniLM-L12-H384-uncased \cite{microsoft/MiniLM-L12-H384-uncased} & MiniLM \\
multi-qa-distilbert-cos-v1 \cite{multi-qa-distilbert-cos-v1} & distilbert-base \cite{distilbert-base} & DBERT \\
\end{tabular}
\label{tab:model_performance}
\end{table}

\section{Performance Evaluation}
To evaluate the ranking performance of the Sentence Transformers model compared to FastText, we adopt the Normalized Discounted Cumulative Gain (nDCG), a standard measure in information retrieval that accounts for both the relevance and the position of retrieved items \cite{10.1145/582415.582418,wang2013theoretical}.

Each retrieved item $r$  is associated with a relevance score $g(r)$. Although relevance scores are defined in the range [0,100], in our setting each control is associated with a limited number of relevant metrics, and evaluation effectively reduces to a binary relevance scenario.

Given a ranked list $R = [r_1, r_2, \ldots, r_k]$, the Discounted Cumulative Gain at position ( k ) is defined as:
\begin{equation}
DCG_k(R) = \sum_{i=1}^k \frac{g(r_i)}{\log_2(i+1)}
\end{equation}

The Ideal Discounted Cumulative Gain $IDCG_k$ is computed over the ideal ranking $R^*$, where relevant items are ranked at the top:
\begin{equation}
IDCG_k(R^*) = \sum_{i=1}^k \frac{g(r_i^*)}{\log_2(i+1)}
\end{equation}

The normalized DCG is then defined as:
\begin{equation}
nDCG_k(R) = \frac{DCG_k(R)}{IDCG_k(R^*)}
\end{equation}

The nDCG score ranges between 0 and 1, where 1 indicates a perfect ranking.

We use $k = 10$. A metric found in the top-10 increases the score. The increase is discounted logarithmically by rank, so top positions weigh more. If no metric is found, the score is zero.

The experiments are conducted considering two variants of nDCG:
\begin{itemize}
\item \textbf{nDCG@10}: We compute nDCG@10 across all 70 controls to evaluate the overall effectiveness of the retrieval system. This metric assigns a score of 0 when no relevant metric is retrieved within the top-10 results, thus jointly capturing both retrieval success and ranking quality.
\item \textbf{Conditional nDCG@10}: To isolate ranking performance, we additionally compute nDCG@10 only over the subset of queries for which at least one relevant metric is retrieved within the top-10 results. This conditional measure evaluates how well the model ranks the correct metric given that it is successfully retrieved.
\end{itemize}

For illustrative purposes, in Table \ref{tab:ndcg_example}, we consider a simplified setting where each control is associated with a single relevant metric. In this case, the relevance is effectively binary, allowing us to clearly highlight the difference between standard nDCG@10 and its conditional variant.

\begin{table}[htbp]
\centering
\caption{Comparison of nDCG@10 and Conditional nDCG@10}
\resizebox{\columnwidth}{!}{%
\begin{tabular}{l c c c}
\toprule
\textbf{Metric} & \textbf{Rank} & \textbf{nDCG@10} & \textbf{Included?} \\
\midrule
Metric A & 1 & 1.00 & Yes \\
Metric B & 3 & 0.50 & Yes \\
Metric C & $>10$ & 0.00 & No \\
\midrule
\multicolumn{2}{l}{\textbf{nDCG@10 Avg}} & \multicolumn{2}{c}{$\frac{1.00 + 0.50 + 0.00}{3} = \mathbf{0.50}$} \\[4pt]
\multicolumn{2}{l}{\textbf{Conditional nDCG@10 Avg}} & \multicolumn{2}{c}{$\frac{1.00 + 0.50}{2} = \mathbf{0.75}$} \\
\bottomrule
\end{tabular}%
}
\vspace{0.2cm}
\label{tab:ndcg_example}
\end{table}

The gap between the two averages highlights the impact of retrieval failures: while standard nDCG@10 penalizes missing relevant items, the conditional variant isolates ranking quality when at least one relevant item is retrieved.

\subsection{Results}\label{sec:results}

\begin{figure}[htbp]
  \centering
  \includegraphics[width=0.5\textwidth]{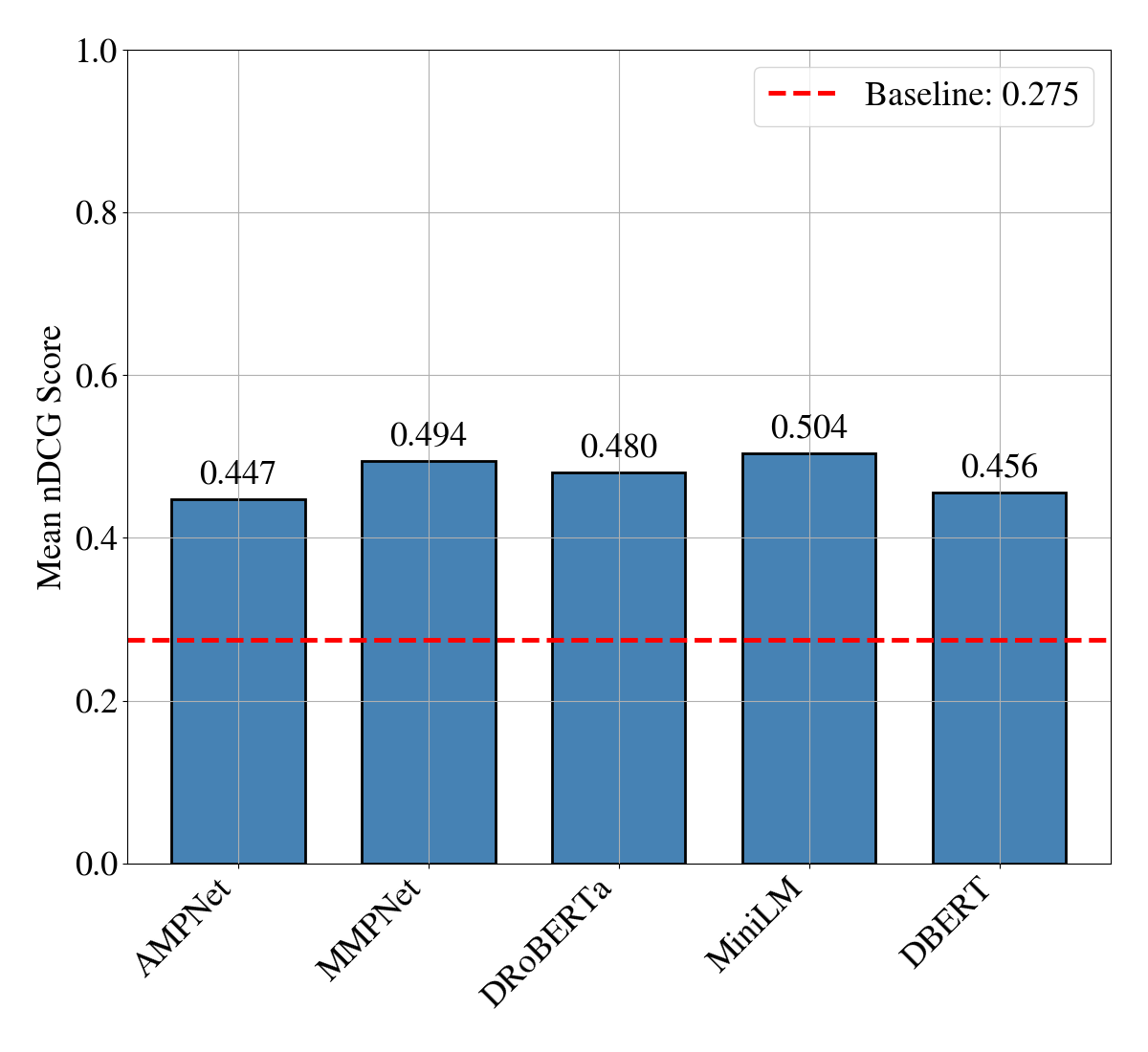}
  \caption{nDCG@10 scores averaged over all controls (including zero scores).\label{img:resultsNdcgScoresForAll}}
\end{figure}

Figure \ref{img:resultsNdcgScoresForAll} reports the standard nDCG@10 scores, computed across all controls, including cases where no relevant metric is retrieved within the top-10 results (i.e., zero scores). This measure captures both retrieval success and ranking quality. The best-performing model, \texttt{all-MiniLM-L12-v2}, achieves an nDCG@10 score of 0.504, compared to the FastText baseline of 0.275.

\begin{figure}[htbp]
  \centering
  \includegraphics[width=0.5\textwidth]{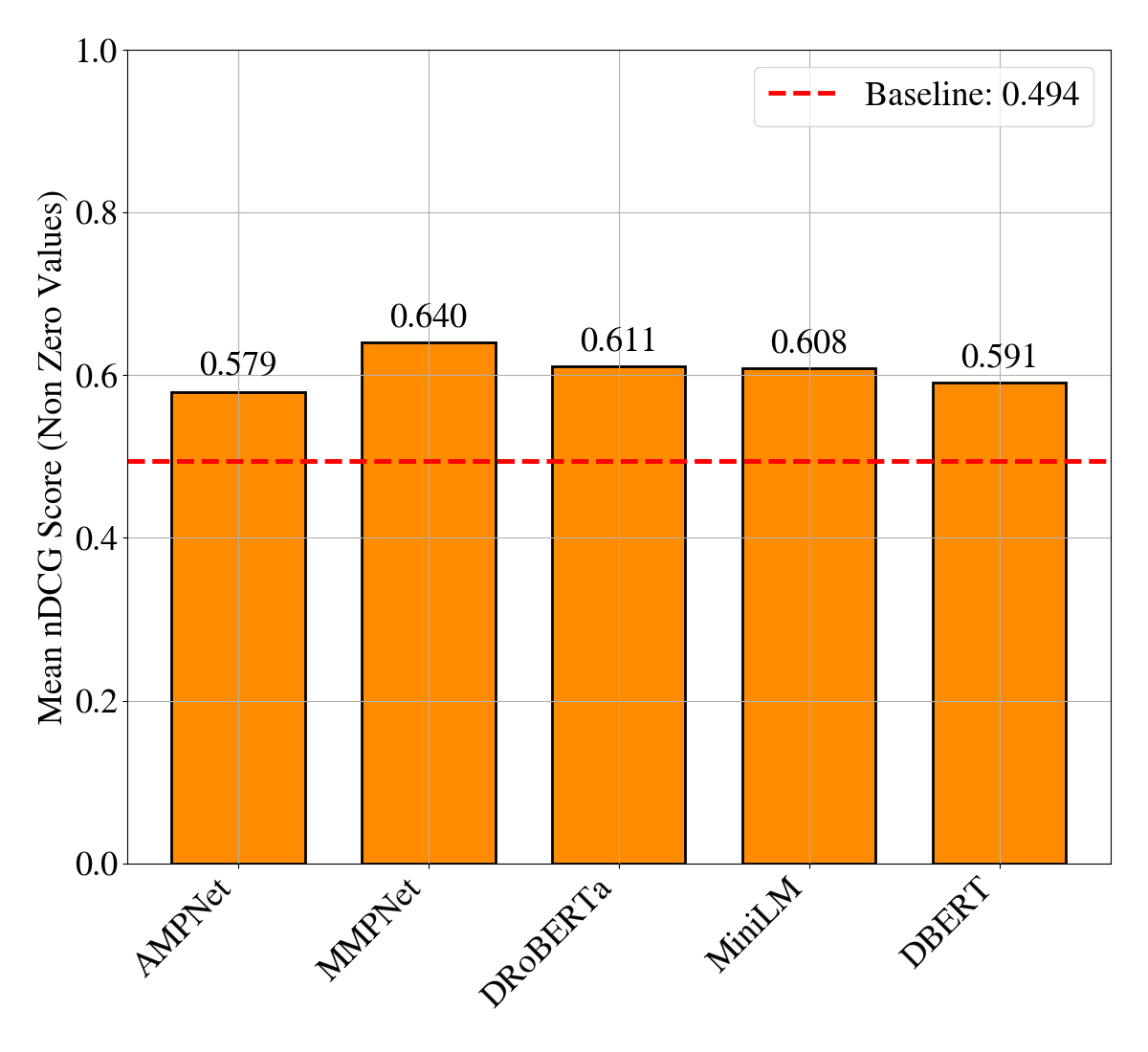}
  \caption{Conditional nDCG@10 scores averaged over controls with at least one relevant result in the top-10.\label{img:resultsNdcgScoresForNonZeroes}}
\end{figure}

Figure \ref{img:resultsNdcgScoresForNonZeroes} further reports the conditional nDCG@10 scores, computed only over the subset of controls for which at least one relevant metric is retrieved within the top-10 results. This measure isolates ranking quality given successful retrieval. \texttt{multi-qa-mpnet-base-dot-v1} achieves the highest conditional 
nDCG@10 score of 0.640, a noticeable improvement over the baseline of 0.494.

Table \ref{tab:numberOfElements} shows the number of non-zero nDCG@10 values for each model. The models based on Sentence Transformers 
successfully locate the correct metric (54--58 versus 39 for the baseline), meaning they correctly assign metrics to the top 10 positions more often. This corresponds to an improvement of approximately 40–50\% in successful retrievals compared to the baseline.

\begin{table}[!ht]
\caption{Number of controls with at least one relevant metric retrieved in the top 10. This count represents how often a model places at least one correct metric within the top-10 results.}\label{tab:numberOfElements}
\renewcommand{\arraystretch}{1.3} 
    \centering
    \begin{tabular}{lc}
        \textbf{Model name} & \textbf{N° of non-zero nDCG values} \\ \hline
        all-mpnet-base-v2 & 54 \\ 
        multi-qa-mpnet-base-dot-v1 & 54 \\ 
        all-distilroberta-v1 & 55 \\ 
        all-MiniLM-L12-v2 & 58 \\ 
        multi-qa-distilbert-cos-v1 & 54 \\  
        \textbf{Baseline} & \textbf{39} \\ 
    \end{tabular}
\end{table}

\section{Conclusions}
In this work, we evaluated a sentence transformer-based approach for automatically associating cloud security controls with security metrics. The proposed method achieves better performance than the only prior work identified for this task in this domain, which is based on FastText embeddings.

The proposed approach achieves a standard nDCG@10 score of 0.504 and a conditional nDCG@10 score of 0.640, both improving over the FastText-based baseline. These results indicate that sentence transformer models provide more effective semantic representations for the control-metric association task, improving both retrieval success and ranking quality.

Future work will focus on fine-tuning Sentence Transformers on larger and more diverse datasets, including the use of data augmentation techniques (e.g., paraphrasing and synthetic query–metric pairs) to further improve the robustness and accuracy of the system.

\bibliographystyle{IEEEtran}
\bibliography{bibliography}

@IEEEtranBSTCTL{IEEEexample:BSTcontrol,
  CTLdash_repeated_names = "no",
}

@inproceedings{reimers2019sentence,
    title = "Sentence-{BERT}: Sentence Embeddings using {S}iamese {BERT}-Networks",
    author = "Reimers, Nils  and
      Gurevych, Iryna",
    booktitle = "Empirical Methods in Natural Language Processing and Joint Conference on Natural Language Processing (EMNLP-IJCNLP)",
    month = nov,
    year = "2019",
    publisher = "Association for Computational Linguistics",
    doi = "10.18653/v1/D19-1410",
    pages = "3982--3992",
}

@misc{huggingFaceSentenceTransformers,
  howpublished = {\url{https://huggingface.co/sentence-transformers}},
  note = {Online},
  year = {2026},
}

@article{yang2020measurement,
  title={Measurement of semantic textual similarity in clinical texts: comparison of transformer-based models},
  author={Yang, Xi and He, Xing and Zhang, Hansi and Ma, Yinghan and Bian, Jiang and Wu, Yonghui and others},
  journal={JMIR Medical Informatics},
  volume={8},
  number={11},
  pages={e19735},
  year={2020},
  publisher={JMIR Publications Inc., Toronto, Canada}
}

@article{sun2022sentence,
  title={Sentence similarity based on contexts},
  author={Sun, Xiaofei and Meng, Yuxian and Ao, Xiang and Wu, Fei and Zhang, Tianwei and Li, Jiwei and Fan, Chun},
  journal={Transactions of the Association for Computational Linguistics},
  volume={10},
  pages={573--588},
  year={2022},
  publisher={MIT Press One Broadway, 12th Floor, Cambridge, Massachusetts 02142, USA~…}
}

@article{DBLP:journals/corr/abs-1810-04805,
  author       = {Jacob Devlin and
                  Ming{-}Wei Chang and
                  Kenton Lee and
                  Kristina Toutanova},
  title        = {{BERT:} Pre-training of Deep Bidirectional Transformers for Language
                  Understanding},
  journal      = {CoRR},
  volume       = {abs/1810.04805},
  year         = {2018},
}

@inproceedings{chen2022semeval,
  author       = {Xi Chen and
                  others},
  title        = {SemEval-2022 Task 8: Multilingual news article similarity},
  booktitle    = {16th International Workshop on Semantic Evaluation, SemEval@NAACL 2022},
  pages        = {1094--1106},
  publisher    = {Association for Computational Linguistics},
  year         = {2022},
  url          = {https://doi.org/10.18653/v1/2022.semeval-1.155},
  doi          = {10.18653/V1/2022.SEMEVAL-1.155},
}

@misc{sentenceTransformersChart,
    author = "{UKPLab}",
    title = {{Pretrained Models}},
    howpublished = {\url{https://www.sbert.net/docs/sentence_transformer/pretrained_models.html}},
    note = {Online} ,
    year=2026,
}

@techreport{medinad2,
  author       = "{MEDINA Consortium}",
  title        = "{Deliverable 2.2 - Continuously certifiable technical and organizational measures and catalogue of cloud security metrics-v2}",
  year         = {2023},
  institution = "European Union’s Horizon"
}

@misc{ENISA2020,
  author       = {ENISA},
  title        = {{EUCS – Cloud Services Scheme}},
  year         = {2020},
  howpublished = {\url{https://www.enisa.europa.eu/publications/eucs-cloud-service-scheme}}
}

@InProceedings{wang2013theoretical,
  title = 	 {A Theoretical Analysis of {NDCG} Type Ranking Measures},
  author = 	 {Wang, Yining and Wang, Liwei and Li, Yuanzhi and He, Di and Liu, Tie-Yan},
  booktitle = 	 {Proceedings of the 26th Annual Conference on Learning Theory},
  pages = 	 {25--54},
  year = 	 {2013},
  volume = 	 {30},
  month = 	 {12--14 Jun},
  publisher =    {PMLR},
}

@article{10.1145/582415.582418,
author = {J\"{a}rvelin, Kalervo and Kek\"{a}l\"{a}inen, Jaana},
title = {Cumulated gain-based evaluation of {IR} techniques},
year = {2002},
issue_date = {October 2002},
publisher = {Association for Computing Machinery},
volume = {20},
number = {4},
issn = {1046-8188},
doi = {10.1145/582415.582418},
journal = {ACM Trans. Inf. Syst.},
month = {oct},
pages = {422–446},
numpages = {25}
}

@misc{EUCSA, 
author= {{European Commission}},
title= {{Regulation (EU) 2019/881 of the European Parliament and of
the Council of 17 April 2019 on ENISA (the European Union Agency for Cybersecurity) and
on information and communications technology cybersecurity certification and repealing
Regulation (EU) No 52}},
      year ={2019} ,
      howpublished ={\url {https://eurlex.europa.eu/eli/reg/2019/881/oj}}}

@misc{CSPcert, 
author= {{CSPCERT Working Group}},
title= {Recommendations for the implementation of the {CSP}
Certification scheme},
year ={2019}
}

@online{BSIc5,
  author = {{Federal Office for Information Security (BSI)}},
  title  = {Cloud Computing Compliance Criteria Catalogue (C5)},
  year   = {2020},
  url    = {https://www.bsi.bund.de/EN/Themen/Unternehmen-und-Organisationen/Informationen-und-Empfehlungen/Empfehlungen-nach-Angriffszielen/Cloud-Computing/Kriterienkatalog-C5/kriterienkatalog-c5_node.html}
}

@online{secnumcloud,
  author = {{ANSSI}},
  title  = {Referentiels d'exigences pour la qualification},
  year   = {2025},
  url    = {https://cyber.gouv.fr/offre-de-service/solutions-certifiees-et-qualifiees/comprendre-levaluation-de-securite/qualification-de-produit-et-services/referentiels-qualification/}
}

@misc{all-mpnet-base-v2,
    howpublished = {\url{https://huggingface.co/sentence-transformers/all-mpnet-base-v2}},
    note = {Online},
    year = 2026,
}

@misc{multi-qa-mpnet-base-dot-v1,
    howpublished = {\url{https://huggingface.co/sentence-transformers/multi-qa-mpnet-base-dot-v1}},
    note = {Online} ,
    year=2026,
}

@misc{all-distilroberta-v1,
    howpublished = {\url{https://huggingface.co/sentence-transformers/all-distilroberta-v1}},
    note = {Online} ,
    year=2026,
}

@misc{all-MiniLM-L12-v2,
    howpublished = {\url{https://huggingface.co/sentence-transformers/all-MiniLM-L12-v2}},
    note = {Online} ,
    year=2026,
}

@misc{multi-qa-distilbert-cos-v1,
    howpublished = {\url{https://huggingface.co/sentence-transformers/multi-qa-distilbert-cos-v1}},
    note = {Online} ,
    year=2026,
}

@misc{microsoft/mpnet-base,
    howpublished = {\url{https://huggingface.co/microsoft/mpnet-base}},
    note = {Online} ,
    year=2026,
}

@misc{distilroberta-base,
    howpublished = {\url{https://huggingface.co/sentence-transformers/all-distilroberta-v1}},
    note = {Online} ,
    year=2026,
}

@misc{microsoft/MiniLM-L12-H384-uncased ,
    howpublished = {\url{https://huggingface.co/microsoft/MiniLM-L12-H384-uncased}},
    note = {Online} ,
    year=2026,
}

@misc{distilbert-base,
    howpublished = {\url{https://huggingface.co/sentence-transformers/stsb-distilbert-base}},
    note = {Online} ,
    year=2026,
}

\end{document}